\documentstyle[twoside,fleqn,espcrc2]{article}
 
\begin{document}

\def\a{\alpha}
\def\b{\beta}
\def\c{\gamma}
\def\d{\delta}
\def\e{\epsilon}
\def\h{\eta}
\def\k{\kappa}
\def\l{\lambda}
\def\m{\mu}
\def\n{\nu}
\def\o{\theta}
\def\p{\pi}
\def\r{\rho}
\def\s{\sigma}
\def\t{\tau}
\def\u{\upsilon}
\def\w{\omega}
\def\x{\chi}
\def\y{xi}
\def\z{\zeta}
 
\def\C{\Gamma}
\def\D{\Delta}
\def\L{\Lambda}
\def\O{\Theta}
\def\P{\Pi}
\def\S{\Sigma}
\def\W{\Omega}

\def\pl{\partial}
\def\rta{\rightarrow}
\def\la{\langle}
\def\ra{\rangle}

\newcommand{\beq}{\begin{equation}}
\newcommand{\eeq}{\end{equation}}
\newcommand{\beqa}{\begin{eqnarray}}
\newcommand{\eeqa}{\end{eqnarray}}

\newcommand{\ttbs}{\char'134}
\newcommand{\AmS}{{\protect\the\textfont2
  A\kern-.1667em\lower.5ex\hbox{M}\kern-.125emS}}

\title{$\eta'\rightarrow\gamma\gamma$ and the topological susceptibility}

\author{G.M. Shore \address{TH Division, CERN\\
        CH1211 Geneva 23, Switzerland.}
\thanks{Permanent address: Dept.~of Physics, University of Wales
	Swansea, Singleton Park, Swansea, SA2 8PP, U.K. }
\thanks{CERN--TH/99-238, SWAT/99-235, hep-ph/9908273. To appear in the 
Proceedings, QCD99 Montpellier} }

\begin{abstract}
The radiative decays $\eta'(\eta)\rta\gamma\gamma$ are discussed. 
The modifications of the conventional PCAC formulae due to the gluonic 
contribution to the flavour singlet axial anomaly are given.
The decay constants satisfy a modified Dashen formula
which generalises the Witten--Veneziano formula for the mass of the $\eta'$.
It is shown how the topological susceptibility
in QCD with massive, dynamical quarks may be extracted from measurements
of $\eta'(\eta)\rta\gamma\gamma$.
\end{abstract}
 
\maketitle
 
\section{The Result}

Radiative decays of the $\eta'$ are currently attracting renewed interest.
(See \cite{Seta} for references.)
In this talk we summarise the special features that arise 
due to the gluonic contribution to the $U_A(1)$ anomaly when PCAC
methods (including chiral Lagrangians) are used in the flavour singlet
channel. In \cite{SVeta}, we presented
an analysis of $\eta' \rta \c\c$ decay in the chiral limit
of QCD, taking into account the gluonic anomaly and the associated anomalous 
scaling implied by the renormalisation group. Here, we summarise the results
of a recent new analysis\cite{Seta} extending this 
to QCD with massive quarks, incorporating $\eta - \eta'$ mixing. In
particular, we show how a combination of the radiative decay formula
and a generalisation of the Witten--Veneziano mass formula for the $\eta'$ 
may be used to measure the gluon topological
susceptibility $\chi(0)$ in full QCD with massive quarks.

Our main result is summarised in the formulae:
\beq
f^{a\a}~ g_{\eta^\a\c\c} ~+~ 2n_f A~ g_{G\c\c}~ \d_{a0} ~=~ 
a_{\rm em}^a {\a\over\pi}
\eeq
which describes the radiative decays, and
\beqa
&f^{a\a} (m^2)_{\a\b} f^{T\b b} = (2n_f)^2 A~\d_{a0}\d_{b0}\cr  
&-2 d_{abc} {\rm tr}~ T^c
\left(\matrix{m_u\la\bar u u\ra &0 &0 \cr
0 &m_d\la\bar d d\ra &0 \cr 
0 &0 &m_s\la\bar s s\ra } \right) \cr
\eeqa
which defines the decay constants appearing in (1) through a 
modification of Dashen's formula to include the gluon contribution to 
the $U_A(1)$ anomaly. 

In these formulae, $\eta^\a$ denotes the neutral pseudoscalars 
$\pi^0,\eta,\eta'$. The (diagonal) mass matrix is $(m^2)_{\a\b}$ 
and $g_{{\eta^\a}\c\c}$ is the appropriate coupling, 
defined as usual from the decay amplitude by
$\la \c\c |\eta^\a \ra = -i g_{\eta^\a\c\c} \e_{\l\r\a\b}p_1^\a p_2^\b
\e^\l(p_1) \e^\r(p_2)$. 
The constant $a_{\rm em}^a$ is the coefficient of the
electromagnetic contribution to the axial current anomaly:
\beq
\pl^\m J_{\m5}^a = d_{acb} m^c \phi_5^b  +
2n_f \d_{a0} Q  + a_{\rm em}^a {\a\over8\pi} F^{\m\n}\tilde F_{\m\n} 
\eeq
Here, $J_{\m5}^a$ is the axial current, $\phi_5^a = \bar q \c_5 T^a q$
is the quark pseudoscalar and $Q = {\a_s\over8\pi}{\rm tr}G^{\m\n} 
\tilde G_{\m\n}$ is the gluon topological charge. 
$m^a$ are the quark masses (see eq(11)). $a=0,3,8$ is the flavour 
index, $T^{3,8}$ are $SU(3)$ generators and $T^0 = {\bf 1}$.
The $d$-symbols are defined by $\{T^a,T^b\} = d_{abc} T^c$.

The decay constants $f^{a\a}$ in (1) are {\it defined} by the relation (2).
In general they are {\it not} the couplings of the pseudoscalar mesons 
to the axial current\cite{SVeta}. In the flavour singlet 
sector, such a definition would give a RG non-invariant decay constant which 
would not coincide with the quantities arising in the correct decay 
formula (1). In contrast, {\it all} the quantities in the formulae
(1),(2) are separately RG invariant\cite{Seta},\cite{Szuoz}. The proof is not 
immediately obvious, and depends on the RGEs for the various Green functions
and vertices defining the terms in (1),(2) being evaluated on-shell or at
zero-momentum.

In practice, since flavour $SU(2)$ symmetry is almost exact, the relations
for $\pi^0$ decouple and are simply the standard ones with $f^{3\pi}$
identified as $f_\pi$ (see eqs(24),(27)).
In the octet-singlet sector, however, there is mixing and the decay constants
form a $2\times 2$ matrix:
\beq
f^{a\a} = \left(\matrix{f^{0\eta'} & f^{0\eta} \cr
f^{8\eta'} & f^{8\eta}} \right) 
\eeq
The four components are independent. In particular, for broken $SU(3)$,
there is no reason to express $f^{a\a}$ as a diagonal matrix times an
orthogonal $\eta - \eta'$ mixing matrix, which would give just three
parameters. Several convenient parametrisations may be made, 
e.g.~involving two constants and two mixing angles, but this does not seem 
to reflect any special dynamics. 

The novelty of our results of course lies in the extra terms arising in
(1) and (2) due to the gluonic contribution to the $U_A(1)$ anomaly.
The coefficient $A$ is the non-perturbative number which specifies the 
topological susceptibility in full QCD with massive dynamical quarks.
The topological susceptibility is defined as
\beq
\chi(0) = \int d^4x~i\la0|T~Q(x)~Q(0)|0\ra
\eeq
The anomalous chiral Ward identities determine 
its dependence on the quark masses and condensates up to an 
undetermined parameter, viz.
\beq
\chi(0) = -A \biggl(1 - A \sum_q {1\over m_q \la \bar q q\ra}\biggr)^{-1} 
\eeq
Notice how this satisfies the well-known result that $\chi(0)$ vanishes
if any quark mass is set to zero.

The modified flavour singlet Dashen formula is in fact a generalisation 
of the Witten--Veneziano mass formula for the $\eta'$. 
Here, however, we do {\it not} impose the leading order in $1/N_c$ 
approximation that produces the Witten--Veneziano formula. 
Recall that this states
\beq
m_{\eta'}^2 + m_{\eta}^2 - 2 m_K^2 =    
-{6\over f_\pi^2} \chi(0)\big|_{\rm YM}
\eeq
To recover (7) from our result (see the first of eqs(9)) the condensate 
$m_s \la \bar s s\ra$ is replaced by the term proportional to $f_\pi^2 m_K^2$ 
using a standard Dashen equation, and the singlet decay constants are set to 
$\sqrt{2n_f}f_\pi$.
The identification of the large $N_c$ limit of the coefficient $A$ with
the non-zero topological susceptibility of pure Yang-Mills theory follows
from large $N_c$ counting rules and is explained in ref\cite{Seta}.

The final element in (1) is the extra `coupling' $g_{G\c\c}$
in the flavour singlet decay formula, which arises because even in the 
chiral limit the $\eta'$ is not a Goldstone boson because of the gluonic
$U_A(1)$ anomaly. A priori, this is {\it not} a physical coupling,
although (suitably normalised) it could be modelled as the coupling of the 
lightest predominantly glueball state mixing with $\eta'$. However,
this interpretation would probably stretch the basic dynamical 
assumptions underlying (1) too far, and is not necessary
either in deriving or interpreting the formula. In fact, the $g_{G\c\c}$
term arises simply because in addition to the electromagnetic anomaly
the divergence of the axial current contains both the quark pseudoscalar  
$\phi_5^a$ and the gluonic anomaly $Q$. 
Diagonalising the propagator matrix for these operators isolates the
$\eta$ and $\eta'$ poles, whose couplings to $\c\c$ give the usual
terms $g_{\eta\c\c}$ and $g_{\eta'\c\c}$. However, the remaining
operator (which we call $G$) also couples
to $\c\c$ and therefore also contributes to the decay formula, whether
or not we assume that its propagator is dominated by a `glueball' pole.

Of course, the presence of the coupling $g_{G\c\c}$ in (1) appears to 
remove any predictivity from the $\eta'\rta\c\c$ decay formula.
In a strict sense this is true, but we shall argue that it may
nevertheless be a good dynamical approximation to assume $g_{G\c\c}$ is 
small compared to $g_{\eta'\c\c}$. In this case, we can combine eqs (1)
and (2) to give a measurement of the non-perturbative coefficient $A$ in
$\chi(0)$. To see this, assume $m_u=m_d\simeq 0$. The terms in (1), (2)
involving the pion then decouple leaving the following five equations,
in which we assume the physical quantities $m_{\eta}, m_{\eta'}, 
g_{\eta\c\c}$ and $g_{\eta'\c\c}$ are all known and we neglect $g_{G\c\c}$.
The decay equations are:
\beqa
{}&f^{0\eta'} g_{\eta'\c\c} + f^{0\eta} g_{\eta\c\c}
+ 6 A g_{G\c\c} = a_{\rm em}^0 {\a\over\pi} \cr
{}&f^{8\eta} g_{\eta\c\c} + f^{8\eta'} g_{\eta'\c\c}
= a_{\rm em}^8 {\a\over\pi} \cr
\eeqa
where $a_{\rm em}^0 = {4\over3} N_c$ and $a_{\rm em}^8 = {1\over3\sqrt3} N_c$,
and the Dashen equations are:
\beqa
&\bigl(f^{0\eta'}\bigr)^2 m_{\eta'}^2 +
\bigl(f^{0\eta}\bigr)^2 m_{\eta}^2 = -4m_s \la\bar s s\ra + 36 A \cr
&f^{0\eta'} f^{8\eta'} m_{\eta'}^2 +
f^{0\eta} f^{8\eta} m_{\eta}^2 = {4\over\sqrt3} m_s \la\bar s s\ra \cr
&\bigl(f^{8\eta}\bigr)^2 m_{\eta}^2 +
\bigl(f^{8\eta'}\bigr)^2 m_{\eta'}^2 = -{4\over3}m_s \la\bar s s\ra  \cr
\eeqa
Clearly, the two purely octet formulae can be used to find $f^{8\eta}$
and $f^{8\eta'}$ if both $g_{\eta\c\c}$ and $g_{\eta'\c\c}$ are known.
The off-diagonal Dashen formula then expresses $f^{0\eta}$ in terms of
$f^{0\eta'}$.
This leaves the two purely singlet formulae involving the still-undetermined
decay constant $f^{0\eta'}$, the topological susceptibility coefficient $A$,
and the coupling $g_{G\c\c}$.
The advertised result follows immediately. If we neglect $g_{G\c\c}$,
we can find $f^{0\eta'}$ from the singlet decay formula and thus
determine $A$ from the remaining flavour singlet Dashen formula. 
This is the generalisation of the Witten--Veneziano formula.

Without neglecting $g_{G\c\c}$, the five equations give a self-consistent
description of the radiative decays, but are non-predictive. It is 
therefore important to analyse more carefully whether it is really
legitimate to neglect $g_{G\c\c}.$
The argument is based on the fact that $g_{G\c\c}$ is both OZI suppressed 
{\it and} renormalisation group (RG) invariant\cite{SVeta}. 
Since violations of the OZI rule are
associated with the $U_A(1)$ anomaly, it is a plausible conjecture
that we can identify OZI-violating quantities by their dependence on the 
anomalous dimension associated with the non-trivial renormalisation of 
$J_{\m5}^0$ due to the anomaly. In this way, RG non-invariance can be used 
as a flag to indicate those quantities expected to show large OZI violations.
If this conjecture is correct, then we would expect the OZI rule to be
reasonably good for the RG invariant $g_{G\c\c}$, which would therefore
be suppressed relative to $g_{\eta'\c\c}$.
(An important exception is of course the $\eta'$ mass itself, which
although obviously RG invariant is not zero in the chiral limit 
as it would be in the OZI limit of QCD.) Notice 
that this conjecture has been applied already with some success to the
`proton spin' problem in polarised deep inelastic scattering\cite{Serice}.
A related $1/N_c$ discussion is given in \cite{Seta}.

\section{The Proof}

Consider first QCD by itself without the coupling to electromagnetism.
The axial anomaly is
\beq
\pl^\m J_{\m5}^a = M_{ab}\phi_5^a + 2n_f Q \d_{a0}
\eeq
The notation is defined in ref\cite{Seta}.
The quark mass matrix is written as $m^a T^a$, so  
\beq
\left(\matrix{m_u &0 &0 \cr
0 &m_d &0 \cr
0 &0 &m_s \cr}\right)
= m^0 {\bf 1} + m^3 T^3 + m^8 T^8
\eeq
The condensates are written as 
\beq
\left(\matrix{ \langle \bar u u\rangle &0 &0 \cr 0 &\langle \bar d d\rangle
&0 \cr
0 &0 &\langle \bar s s\rangle \cr}\right) = {1\over 3} \la\phi^0\ra {\bf 1}
+ 2
\la\phi^3\ra T^3 + 2 \la\phi^8\ra T^8 
\eeq
where $\langle \phi^c\rangle$ is the VEV $\langle \bar q T^c q\rangle$.
Then
\beq
M_{ab} = d_{acb} m^c, ~~~~~~
\Phi_{ab} = d_{abc} \langle \phi^c\rangle 
\eeq

The anomalous chiral Ward identities, at zero momentum, for the 
two-point Green functions of these operators are\cite{Seta}
\beqa
&2n_f \la Q~Q\ra \d_{a0} + M_{ac} \la \phi_5^c~Q\ra = 0 \cr
&2n_f \la Q~\phi_5^b\ra \d_{a0} + M_{ac}\la\phi_5^c~\phi_5^b\ra 
+ \Phi_{ab} = 0 \cr
\eeqa
which imply
\beq
M_{ac} M_{bd} \la \phi_5^c~\phi_5^d\ra = - (M\Phi)_{ab} 
+ (2n_f)^2 \la Q~Q\ra \d_{a0}\d_{b0}
\eeq
We also need the result for the general form of the topological 
susceptibility (see eq(6)):
\beq
\chi(0) \equiv \la Q~Q \ra = {-A\over 1 - (2n_f)^2 A (M\Phi)_{00}^{-1} }
\eeq

Although the pseudoscalar operators $\phi_5^a$ and $Q$ indeed couple to the 
physical states $\eta^\a = \eta', \eta, \pi^0$, it is more convenient to
redefine linear combinations such that the resulting propagator matrix
is diagonal and properly normalised. So we define operators
$\eta^\a$ and $G$ such that
\beq
\left(\matrix{\la Q~Q \ra &\la Q~\phi_5^b \ra \cr
\la \phi_5^a ~Q \ra &\la \phi_5^a~\phi_5^b \ra }\right) ~~\rta~~
\left(\matrix{\la G~G\ra &0 \cr
0 &\la \eta^\a ~ \eta^\b \ra }\right)
\eeq
This is achieved by 
\beqa
G &=  Q - \la Q~\phi_5^a\ra (\la \phi_5 ~\phi_5\ra)_{ab}^{-1} \phi_5^b \cr
&= Q + 2n_f A \Phi_{0b}^{-1} \phi_5^b ~~~~~~~~~~~~\cr
\eeqa
\vfill\eject
\noindent and
\beq
\eta^\a = f^{T\a a} \Phi_{ab}^{-1} \phi_5^b 
\eeq
With this choice, the $\la G~G\ra$ propagator is
\beq
\la G~G\ra = - A
\eeq
and we impose the normalisation
\beq
\la \eta^\a ~\eta^\b\ra = {-1\over k^2 - m_{\eta^\a}^2}\d^{\a\b}
\eeq
This implies that the constants $f^{a\a}$ in (19), which 
are simply the decay constants, must satisfy the (Dashen) identity
\beqa
f^{a\a} m_{\a\b}^2 f^{T\b b} &= \Phi_{ac} (\la \phi_5~\phi_5 \ra)_{cd}^{-1}
\Phi_{db} ~~~~~~~~~~ \cr
&= -(M\Phi)_{ab} + (2n_f)^2 A \d_{a0}\d_{b0} \cr
\eeqa
The last line follows from the Ward identities (15) and (16) .
In terms of these new operators, the anomaly equation (10) is:
\beq
\pl^\m J_{\m5}^a = f^{a\a} m_{\a\b}^2 \eta^\b + 2n_f G \d_{a0}
\eeq

Now recall how conventional PCAC is
applied to the calculation of $\pi^0\rta\c\c$. The pion decay constant
is defined as the coupling of the pion to the axial current
\beq
\la 0|J_{\m5}^3|\pi\ra = ik_\m f_\pi ~\Rightarrow~
\la 0|\pl^\m J_{\m5}^3|\pi\ra = f_\pi m_\pi^2
\eeq
and satisfies the usual Dashen formula
The next step is to define a `phenomenological pion field' $\pi$ by
\beq
\pl^\m J_{\m5}^3 \rta f_\pi m_\pi^2 \pi
\eeq
To include electromagnetism, the full anomaly equation is extended as in
(3) to include the $F^{\m\n} \tilde F_{\m\n}$ contribution. Using (24)
we have
\beqa
&ik^\m\la \c\c|J_{\m5}^3|0 \ra ~~~~~~~~~~~~~~~~~ \cr
&= f_\pi m_\pi^2 \la \c\c|\pi|0\ra
+ a_{\rm em}^a {\a\over8\pi} \la \c\c|F^{\m\n} 
\tilde F_{\m\n}|0\ra~~~~\cr
&= f_\pi m_\pi^2 \la \pi~\pi\ra \la \c\c|\pi\ra
+ a_{\rm em}^a {\a\over8\pi} \la \c\c|F^{\m\n} \tilde F_{\m\n}|0\ra \cr
\eeqa
where $\la \pi~\pi\ra$ is the pion propagator $-1/(k^2-m_\pi^2)$.
At zero momentum, the l.h.s.~vanishes because of the explicit $k_\m$
factor and the absence of massless poles. We therefore find, 
\beq
f_\pi g_{\pi\c\c} = a_{\rm em}^3 {\a\over\pi}
\eeq

In the full theory including the flavour singlet sector and the gluonic
anomaly, we find a similar result. The `phenomenological fields'
are defined by (23) where the decay constants satisfy the generalised
Dashen formula (22) . Notice, however, that they are {\it not} simply
related to the couplings to the axial current as in (24) for the
flavour non-singlet. We therefore find:
\beqa
{}&ik^\m\la \c\c|J_{\m5}^a|0 \ra 
= f^{a\a} m_{\a\b}^2 \la \c\c|\eta^\b|0\ra ~~~~~~~~~~~~~~~~\cr
{}&+ 2n_f \la \c\c|G|0\ra \d_{a0} 
+ a_{\rm em}^a {\a\over8\pi} \la \c\c|F^{\m\n} \tilde F_{\m\n}|0\ra \cr
{}&~~~~~~~~~= f^{a\a} m_{\a\b}^2 \la \eta^\b~\eta^\c \ra \la \c\c|\eta^\c \ra \cr
{}&+ 2n_f \la G~G\ra \la \c\c|G\ra \d_{a0} 
+ a_{\rm em}^a {\a\over8\pi} \la \c\c|F^{\m\n} \tilde F_{\m\n}|0\ra \cr
\eeqa
using the fact that the propagators are diagonal in the basis $\eta^\a, G$.
Using the explicit expressions (20),(21) for the Green functions,
we find in this case:
\beq
f^{a\a} ~g_{\eta^a\c\c} + 2n_f A ~g_{G\c\c} ~\d_{a0} ~=~ a_{\rm em}^a 
{\a\over\pi}
\eeq
where the extra coupling $g_{G\c\c}$ is defined through (28).
This completes the derivation. It is evidently a
straightforward generalisation of conventional PCAC with the 
necessary modification of the usual formulae to take account of the
extra gluonic contribution to the axial anomaly in the flavour
singlet channel, the key point being the identification of the 
operators $\eta^\a$ and $G$ in (23).

Finally, notice that these methods may equally be applied to other decays 
involving the $\eta'$ such as $\eta'\rta V\c$ where $V$ is a light 
$1^-$ meson, $\eta' \rta \pi\pi\c$, $\psi \rta \eta' \c$, etc.

\section{Acknowledgements}

I would like to thank S.~Narison and G.~Veneziano for helpful discussions.
This work was supported by the PPARC grant GR/L56374 and by
the EC TMR grant FMRX-CT96-0008.

\end{document}